\documentclass[preprints,article,accept,moreauthors,pdftex]{Definitions/mdpi}
\firstpage{1} 
\pubvolume{1}
\issuenum{1}
\articlenumber{0}
\pubyear{2022}
\copyrightyear{2022}
\datereceived{} 
\dateaccepted{} 
\datepublished{} 
\hreflink{https://doi.org/} 

\usepackage{xcolor}
\usepackage{caption}
\usepackage{subfig}

\Title{EMAKG: An Enhanced Version Of The Microsoft Academic Knowledge Graph}

\TitleCitation{Title}

\Author{Laura Pollacci $^{1*}$\orcidA{}
}

\AuthorNames{Laura Pollacci}

\AuthorCitation{Pollacci, L.}

\address[1]{%
$^{1}$ Department of Computer Science, University of Pisa; laura.pollacci@di.unipi.it}

\corres{Correspondence: laura.pollacci@di.unipi.it}

\abstract{Scholarly knowledge graphs are valuable sources of information in several research fields. Despite the number of existing datasets related to publications and researchers, resource quality, coverage and accessibility are still limited. This article presents the Enhanced Microsoft Academic Knowledge Graph, a large dataset of information about scientific publications and involved entities, and the methods developed to build it. Data includes geographical information, researchers' collaborative networks and movements between institutions, academic-related metrics, and linguistic features. The dataset merges information from several data sources and has high temporal and spatial coverage, allowing several use cases.}


\dataset{v. 0.0: 10.5281/zenodo.5888647, 
\url{https://github.com/LauraPollacci/EMAKG}}

\datasetlicense{Creative Commons Attribution 4.0 International (CC BY 4.0)}

\begin{document}


\section{Introduction}

\label{sec:introduction}
Sharing knowledge is ever more crucial, especially in scientific research~\cite{ismail2013share}. Data representing highly skilled personnel is a key to interpreting and understanding scientific collaborations and knowledge exchange phenomena. 
Given its multifaceted nature, numerous strands of research are involved in the analysis of highly skilled personnel and scholarly data, including digital libraries \cite{DBLP:conf/semweb/Farber19,farber2019microsoft}, collaborator discovery, expert finding, and recommendation systems~\cite{khan2017survey}. 
Furthermore, scientific networks of collaboration and exchange, understood as physical displacement (mobility), are at the centre of research attention. 
Despite the recent interest in knowledge exchange and the increase in movements of highly-skilled personnel, moving researchers have captured a limited interest.
There exists a notable gap in the understanding of researchers' mobility, knowledge exchange, and scientific collaboration networks, besides a few exceptions 
~\cite{bonisch2016rituals,kim2017academic,koh2020academic}. 
One of the challenges with modelling researchers' mobility and collaborations is the lacking of data and international statistics providing definitions and specific indicators, e.g., socio-economic, educational, and professional indicators \cite{ahmad2021anatomy,willekens2016international}. 
The most recent research has focused on alternative data sources to fill the gaps posed by traditional data, e.g., register statistics. 
Unconventional data describing publications, researchers' careers and movements have opened new research opportunities for multiple fields of study~\cite{Sirbu2020}. 
There is a large variability in terms of available data sources, accessibility, format, coverage, type, and the number of contents, as discussed in Section~\ref{sec:related}. 
Researchers have benefited from alternative data sources to study academic collaborations networks and to develop scientific mobility indicators~\cite{conti2020dynamics,miranda2020scholarly,zhao2020investigation} and to examine the scientific ethnic and mobility networks~\cite{alshebli2018preeminence,vaccario2020mobility}. 

This paper presents the Enhanced MAKG (DOI: 10.5281/zenodo.5888647)\footnote{The composition of the dataset is described in Section \ref{sec:appendice}.}, a large dataset of scientific publications and related entities, including authors, and the methods\footnote{GitHub repository: \url{https://github.com/LauraPollacci/EMAKG}.} developed to build it. 
The proposed dataset originates from the Microsoft Academic Knowledge Graph (MAKG) \cite{DBLP:conf/semweb/Farber19,farberenhancing, farber2019microsoft}, one of the most extensive freely available knowledge graphs on publications. 
I first assess the limitations of the current MAKG dataset in Section~\ref{sec:makg}. 
Then, based on these, several methods are designed to enhance data and facilitate the number of use case scenarios, particularly in mobility and network analysis. 
The dataset provides two main advantages. 
First, it has improved usability, facilitating access to non-expert users. 
Second, it includes an increased number of types of information obtained by integrating various datasets and sources,  which help expand the application domains. For instance, geographical information could help mobility (and migration) research. 
The knowledge graph completeness is improved by retrieving and merging information on publications and other entities no longer available in the latest version of MAKG. 
Furthermore, geographical and collaboration networks details are employed to provide data on authors, including their working connections and movements between institutions and countries, opening several new possible research and use cases for the dataset. 
Further, data is generally enriched and standardised by designing Natural Language Processing (NLP) semi-supervised approaches. 

The rest of this paper is organised as follows. 
Section~\ref{sec:contributions} describes the main contributions in this article. 
Section~\ref{sec:related} discusses available scholarly data resources and their main differences. 
In particular, Section~\ref{sec:makg} provides a general overview of the Microsoft Academic Knowledge Graph with its enhancements over time, current limitations, and usages. 
The methods developed to build the EMAKG are described from Section~\ref{sec:Geolocating_affiliations} to \ref{sec:fos}, while description of the dataset is in Section~\ref{sec:stats}. 
Section~\ref{sec:concl} concludes the paper with the final discussion together and future works, after a brief discussion on possible usages of the Enhanced MAKG in Section~\ref{sec:usage}. 

\section{Contributions}
\label{sec:contributions}

This section summarises the main contributions proposed to enhance and improve the Microsoft Academic Knowledge Graph.

\textbf{Facilitation of Use Case Scenarios}
The dataset dissemination and the use scenarios strictly rely on data accessibility. 
Thus, providing most of the data in Comma Separated Value (.csv) and .txt formats could lower necessary skills to access, manage and analyse them. 
Use cases may also depend on the number, type, and ease of understanding of data. 
To this end, data standardisation is enhanced. 
In addition, entities' properties rely on official coding systems, including \textsc{iso} 3166 codes for countries\footnote{Standard codes for the representation of names of countries and their subdivisions.}, \textsc{issn} for journals\footnote{Eight-digit unique International Standard Serial Number used to identify serial publications, e.g., journals and magazines.}, and \textsc{iso} 639-1 for languages. 

\textbf{Knowledge Graph. }
As discussed in Section~\ref{sec:makg}, the MAKG has often been updated, but such changes also impose a data loss. 
To this end, the knowledge graph completeness is improved by retrieving information no longer available in MAKG from its parent, the Microsoft Academic Graph (MAG)~\cite{sinha2015overview}, e.g., links between institutions, papers, and authors. 
Furthermore, other sources, i.e., Wikipedia, are exploited to add new information on entities. 
For instance, the semi-supervised method in Section~\ref{sec:Geolocating_affiliations} combines $a)$ reverse geocoding, $b)$ information retrieval and $c)$ data integration to provide additional information on affiliations. 
These include homepage, foundation date, type, acronym, and a set of geographical data, e.g., city, country name, and  \textsc{iso} codes. 

\textbf{Mobility. }
As discussed in Section~\ref{sec:makg}, the application scenario of the MAKG lacks mobility-related studies. 
Geographical information and affiliations geolocalisation allow describing researchers' movements over institutions and countries. 
Starting from the hypothesis that an author lives in the country where its affiliation is located, annual publications and geolocalised affiliations are computed by authors. 
The relationship between these provides (a) the \textit{authors' annual location}, as the most frequent country among locations of institutions related to an author's annual publications; and (b) \textit{authors' career's nationality}, as the country of the first geolocated institution over an author's career.
Researchers locations allows dealing with mobility (and migration) related concepts, such as \textit{flows}  and \textit{stocks} (Section~\ref{sec:careers_stocks_flows}). 
Combining the literature \cite{desa2017handbook} and the information on authors' locations and careers, the concept of \textit{i. working-nativity} is introduced. Given an author A and a country C, A is a working-native of C if C corresponds to the author's career's nationality. Furthermore, are defined 
\textit{ii. authors' stocks} the number of authors identified as international working migrants in a given country and year; \textit{iii. authors' flows} the number of authors entering or leaving a given country and year. 
Thus, given a country C, \textit{authors' stocks} are computed by counting the number of non-working-natives authors annually located in C. Moreover, \textit{authors' flows} are modelled as the direct graph between worldwide countries. 

\textbf{Networks of collaborations. }
Similar to mobility, the role and the topology of scientific collaboration networks have been extensively analysed for different research purposes (Section~\ref{sec:networks}). 
Authors' collaboration networks could be employed to understand dynamics between researchers' and study knowledge exchange over institutions and countries. 
According to the literature, authors having authored a paper together are linked on a yearly basis~\cite{newman2001structure}. 
Thus, authors' networks of collaborations are modelled as \textit{ego networks} annually (Section~\ref{sec:networks}). 
Annual authors' ego networks may facilitate network analysis and help understanding dynamics between researchers over time. 

\textbf{Fields of study. }
The latest MAKG version includes a descriptive classification of fields of study. 
Nevertheless, together with the data loss~\cite{farberenhancing}, the newest classification method seems to be more suitable for given research fields instead of others. 
Here, I propose a method that starts from a limited set of top fields of study to propagate them based on parenting relationships. 
The procedure allows obtaining fields of study labelled with one or more top-level disciplines and described by a score in the range $[0,1]$ computed as the \textit{proportion} over the list of inherited labels. 
The obtained labelled fields of study (\textsc{fos}) can help understand phenomena and dynamics about researchers, e.g., exploring trends of publications rates by fields and disciplines' attractiveness (as the number of authors publishing in a given area). 

\textbf{General Enhancements. }
Data is enriched and aggregated by designing Natural Language Processing semi-supervised approaches aiming to include $a)$ academia related metrics (\textit{h-index}); $b)$ abstracts and linguistic features, i.e., standard language codes, \textit{tokens}, and \textit{types}; $c)$ entities' general information, e.g., date of foundation, type and acronym of institutions, among others.


\section{Related Work}
\label{sec:related}

To date, various systems allow exploration of scientific data via repository interfaces and ensure access to integrated datasets from multiple resources~\cite{osborne2013exploring}. 
Major data resources include but are not limited to: Scopus\footnote{Scopus: \url{https://www.scopus.com/.}} a multidisciplinary composite source; 
DBLP\footnote{DBLP: \url{https://dblp.uni-trier.de/}}, a computer science bibliography website; 
Google Scholar\footnote{Google Scholar: \url{https://scholar.google.com/}.}, which allows search and citation services over the academic literature; 
CiteSeerX\footnote{CiteSeerX: \url{http://citeseerx.ist.psu.edu/index}.}, a large-scale harvesting of indexed papers; Web of Science (WOS)\footnote{Web of Science: \url{https://data.mendeley.com/datasets/9rw3vkcfy4/6}.}, a research dataset for scholarly publications; 
Microsoft Academic Search (MAS)\footnote{MAS: \url{http://academic.research.microsoft.com/}.}~\cite{sinha2015overview} which includes co-authorship graphs, publication ranks, and authors' collaborations; 
ArXiv\footnote{ArXiv: \url{https://arxiv.org/help/api/index}.}, which focuses on specific field of studies; 
Springer's  SciGraph\footnote{SciGraph: \url{Seehttps://www.springernature.com/de/researchers/scigraph}.}that include publications from one publisher; 
OpenCitations~\cite{peroni2015setting}, an independent organization for open scholarship that uses Semantic Web (Linked Data) technologies; 
and PubMed/NLM\footnote{PubMed/NLM: \url{https://www.ncbi.nlm.nih.gov/books/NBK3827/}.} that includes life sciences and biomedical contents. 
Moreover, reference management tools, such as Mendeley\footnote{Mendeley: \url{http://www.mendeley.com}.} and Zotero\footnote{Zotero: \url{http://www.zotero.org}.} and researchers-oriented social platforms, such as ResearchGate\footnote{ResearchGate: \url{http://www.researchgate.net}.} and Accademia.edu\footnote{Accademia.edu: \url{http://www.academia.edu}.} can be also used as data resources.
Available resources are heterogeneous in terms of distribution, necessary skills for access, content, and size, and the topic would require a separate discussion. 
However, Google Scholar, MAS, CiteSeerX, and Scopus include multidisciplinary data. On the contrary, ArXiv, DBLP, OpenCitations, and PubMed/NLM are field of study-oriented, and SciGraph is publisher-oriented.
Regarding the size, ArXiv includes just under 2 million articles, Open Citation (Corpus) just over 300 thousand, Google Scholar 170-175 million, and MAS 120 million. 
Finally, MAS, CiteSeerX, DBLP, and Google Scholar include thesis and informal publications, reference works, and books.  
Resources most similar to MAKG include AceKG~\cite{wang2018acekg} a large scale dataset of academic entities, such as papers, authors, fields of study, venues, and institutions; SPedia~\cite{aslam2017spedia} a rich source of bibliographic information about 9 million papers; and the RDF dataset~\cite{gentile2015conference} by Nuzzolese et al.~\cite{nuzzolese2016conference} deriving from the Semantic Web Conference ontology. 
However, AceKG does not provide continuous updates, and the RDF data from Nuzzolese et al. ~\cite{nuzzolese2016conference} is too specific for the purpose of EMAKG since it refers exclusively to Semantic Web conferences.

\subsection{Microsoft Academic Knowledge Graph}
\label{sec:makg}

\begin{table}[t]
    \centering
    \caption{Overview of MAKG (v. 2020-06-19) - "*" refers to disambiguated authors.}
    \begin{tabular}{ll}
        \hline
        \textbf{Subset}	& \textbf{\# entities}\\
        \hline
        Papers & 238,670,900\\
        Authors	& 243,042,675\\
        Authors* & 151,355,324\\
        Affiliations & 25,767\\
        PaperAuthorAffiliations* & 644,154,780\\
        FieldOfStudy & 740,460\\
        \hline
    \end{tabular}
    \label{tab:makg}
\end{table}

The Microsoft Academic Knowledge Graph derives from the Microsoft Academic Graph, an extensive database about scientific publications modelled as a connected knowledge graph. 
The MAKG provides information about scientific publications and entities involved in and related to these, including authors, venues, and institutions. 
The dataset includes data for almost 240 million papers and 245 million authors affiliated to more than 25 thousand institutions, as shown by the distribution of the entities among the main entity types in Table~\ref{tab:makg}. 
The last version of MAKG (v. 2020-06-19\footnote{Data are available at \url{https://doi.org/10.5281/zenodo.4617285}.}) shows notable changes regarding the previous ones such as new properties for entities relations modelling, authors disambiguation, and geographical coordinates for institutions. 
Also, the number of subsets in the dataset changed from 18 to 26. 
The MAKG strongly benefits from the existing resources, such as DBpedia\footnote{DBpedia: \url{https://www.dbpedia.org/}.}, the Dublin Core Metadata Initiative (DCMI)\footnote{Dublin Core Metadata Initiative: \url{https://dublincore.org/}.}, and Semantic Publishing and Referencing (SPAR) ontologies~\cite{peroni2018spar}, which include FaBiO\footnote{FaBiO: the FRBR-aligned Bibliographic Ontology, an ontology for  bibliographic records on the Semantic Web; \url{https://sparontologies.github.io/fabio/current/fabio.html}.}, CiTO\footnote{CiTO: the Citation Typing Ontology, an ontology for the characterization of bibliographic citations; \url{https://sparontologies.github.io/cito/current/cito.html}.}, 
PRISM\footnote{PRISM: the Publishing Requirements for Industry Standard Metadata, \url{https://idealliance.org/specifications/prism-metadata/}.}, DataCite\footnote{DataCite: an ontology that defines identifiers for bibliographic resources and related entities; \url{https://datacite.org/}.}, and C4O\footnote{C4O, 
the Citation Counting and Context Characterization Ontology; \url{https://sparontologies.github.io/c4o/current/c4o.html}.}. 

Due to data richness and high coverage, the MAKG has been employed in several research fields and scenarios, including bibliometrics and scientific impact~\cite{farber2021identifying,schindler2020investigating,tzitzikas2020can}, recommender systems~\cite{kanakia2019scalable},  data analytics (i.e.,  Nesta business intelligence tools\footnote{\url{https://www.nesta.org.uk}.}), and benchmarking~\cite{ajileye2021streaming}. 
Further, the MAG, to which the MAKG originally derives, has been extensively investigated~\cite{herrmannova2016analysis,effendy2017analysing,effendy2016investigations} and used for scientific ethics and mobility networks~\cite{nuzzolese2016conference,gentile2015conference}, and in COVID-19 related studies~\cite{chen2020glimpse,shemilt2021cost}.

Starting with the first version (v. 2018-11-09), the MAKG has notably been enhanced. According with \cite{farberenhancing}, the most significant limitations rely on data replication, the field of study hierarchy, and the scarcity of entity embeddings. 
The authors of the latest MAKG version have managed these issues, making substantial changes to the dataset and its information. 

\begin{itemize}
    \item According to~\cite{wang2020microsoft}, the design of the parent of MAKG leads to more author entities than real authors. 
Thus, in the last version, the replications of author entities have been addressed by performing name disambiguation. 
    \item Originally, the MAKG structures the information on research areas as its parent. Fields of study are organised in a multi-level hierarchy where parent research areas are fine-grained and multiple. The last MAKG version provides a descriptive classification of fields of study based on abstracts of publications. 
    \item In previous versions, entity pre-trained embeddings are provided only for publications using RDF2Vec. The last MAKG version includes embeddings for journals, conferences, and authors. 
\end{itemize}

Along with these improvements, it is also worth noting the inclusion of the geographic coordinates of the institutions, i.e., affiliations.

One of the strengths of the MAKG is its free distribution, which allows free, direct and unlimited access via RDF knowledge graph dump with resolvable URIs\footnote{\url{https://makg.org}.}, a public SPARQL endpoint\footnote{\url{ http://ma-graph.org/sparql}.} and via Zenodo\footnote{\url{http://doi.org/10.5281/zenodo.4617285}.}.
Although the advantages of the triples-RDF format are pointed out in~\cite{farber2019microsoft}, the skill level needed to access the dataset could represent a limit.
On the one hand, the MAKG has already proven to be a highly versatile resource across both study scenarios and research fields. 
On the other hand, the RDF format may represent a limit - or at least - a challenge for some researchers, students and non-professionals.
By providing .csv and .txt data, even users with fewer computer skills can easily access avoiding software and online platforms for format conversion, even given the size.
In addition, splitting the dataset allows users to download, manage and store subsets individually, as MAKG and MAG.
The application scenario of the MAKG lacks mobility-related studies. 
Conversely, the MAG has been already exploited to analyse scientific ethics and mobility networks~\cite{nuzzolese2016conference, gentile2015conference}. 
This could be due to (a) the scarcity of geographic information and (b) the lack of standardised geographic data. 
The latest MAKG version only partially adds geographic information, i.e., coordinates (longitude and latitude) for affiliations. 
However, geographic data are not standardised, e.g., the data on conferences includes the DBpedia location (city). 
The country is not provided, and several homonymous cities exist\footnote{E.g., London is in the United Kingdom and Ontario, Milan is in Italy and Ohio, and Paris is in France, Texas, and Tennessee.}. 
As for the data format, standardised and easy-accessible geographical information could favour multidisciplinary mobility-related studies.
Others limitation of the MAKG depends on its design. 
First, the MAKG provides only the last affiliation of authors~\cite{farber2019microsoft}, making it impossible to build geolocalised authors' careers. 
On the contrary, the MAG provides the relationship between papers, authors and affiliations, allowing re-integrating data useful to locate authors over time. 
Secondly, the new design choice to label papers with fields of study based on the abstracts (which are not available for all papers) imposes a loss of data~\cite{farberenhancing}. 
Also, the results show that the design is more suitable for some research fields. 
The performance could suffer from the low specificity of abstracts' terms since the best performances are for fields with strictly domain-dependent vocabularies~\cite{farberenhancing}. 


\section{Geolocating and Enriching Affiliations}
\label{sec:Geolocating_affiliations}

Geographic coordinates, i.e., latitude and longitude, for affiliations is one of the improvements of the latest MAKG version. 
However, enhancing the geographical dimension of the dataset could facilitate studies in human mobility and migration, especially of scholars. 
In the latest decades, researchers have benefited from advantages gained from alternative data sources such as bibliometric repositories, e.g., Scopus and Web of Science, to study academic collaboration networks and scientific mobility indicators~\cite{conti2020dynamics,miranda2020scholarly,zhao2020investigation}. 
However, international scientific mobility and migration patterns are still not fully explored besides a few studies ~\cite{moed2014bibliometric,robinson2019many}. 
Thus, geolocating affiliations aim to add standardised and detailed metadata facilitating mobility-related studies.

\subsection{Reverse Geocoding}
\label{subsec:reverse_geocoding}

By leveraging coordinates provided in MAKG, reverse geocoding methods can be applied to transform $(latitude, longitude)$ pairs into addresses - or at least - parts of it, e.g., country name. 
To this end, an ad-hoc semi-supervised NLP function is built to return an array of standardised geographical metadata, including the city name, state, postcode, the country with its \textsc{iso} 3361 codes and official name, from coordinates. 
The algorithm takes as inputs the affiliations' latitude and longitude coordinates and applies Geopy\footnote{\url{https://geopy.readthedocs.io/en/stable/}.} and Reverse Geocoder\footnote{Reverse Geocoder: \url{https://pypi.org/project/reverse_geocoder/}.} reverse geocoding methods. 
Then, results are cross-checked following a set of rules to assign at least a country - together with standardised geographical information - to each affiliation. 
Geopy and Reverse Geocoder results show that the libraries classify some countries differently, e.g., Unincorporated territories of the United States\footnote{E.g., Puerto Rico, American Samoa, United States Virgin Islands, which are labelled with their Alpha 2 country code (PR, AS, VI) or with the US code depending on the library.}. 
Since both Alpha 2 codes are correct depending on the country classification, i.e., including Unincorporated territories of the United States under the US label or not since non-incorporated countries, EMAKG provides both. 
Once all affiliations are labelled with an Alpha 2 code, the algorithm uses the PyCountry\footnote{\url{https://github.com/flyingcircusio/pycountry}.} library to retrieve Alpha 3 \textsc{iso} 3361 code and official name 
plus the country name.

\subsection{\textsc{url}-based pipeline}
\label{subsec:url-based}
An ad-hoc pipeline provides various information from Wikipedia \textsc{url}s, including geographic data. 
This algorithm uses a parser based on WpTools\footnote{WpTools: \url{https://wp-tools.com/author/wptools/}.} to obtain an array of raw information from Wikipedia \textit{infoboxes} given a \textsc{url}.
The parser provides geographical and non-geographical data, such as country, state, city, acronym, foundation date, and homepage. 
However, city, homepage, and foundation date labels are inconsistent between affiliations thus keyword sets are used to gather data\footnote{I.e., city: \textit{[city, location, headquarter],} foundation\_date = \textit{foundation, foundation\_date, established]}, homepage = \textit{[homepage, \textsc{url}, website]}.}.  
For non-geographical fields, supervised NLP-based rules allow extracting and standardising valuable data.
Standardised geographical information, i.e., city name with coordinates (latitude and longitude), state, a country name with \textsc{iso} 3361 Alpha 2 and Alpha 3 codes and official name are provided by a two-step \textsc{url}-based geolocation algorithm. 
First, the algorithm uses city labels to search for city names in GeoText cities set. 
It then retrieves country names and related data such as country \textsc{iso} 3361 alpha codes and official names by applying a support function. This uses the raw country text to retrieve the standardised and official name with \textsc{iso} 3361 alpha codes from the GeoText country set. 
While the ``fuzzy search'' provided by PyCountry is used if GeoText does not give results. 
Also, the support function includes two supplementary methods used if a country is still not found. 
The first evaluates whether the state field can represent a valid country by re-applying the main support function with the state as the input parameter. 
The second one extracts the country from the city name with Geopy and gathers related data re-applying the main support function. 

\subsection{Enriching Affiliations}
\label{subsec:enrich_affiliations}

Both reverse geocoding (Section \ref{subsec:reverse_geocoding}) and \textsc{url}-based geolocation (Section \ref{subsec:url-based}) algorithms are applied to all affiliations, and results are cross-checked and integrated by a set of semi-supervised NLP-based rules. 
The method iterates over all the reversed geocoded affiliations, checking if coordinates and country are retrieved using reverse geocoding. If not, geographical data extracted \textsc{url}-based algorithm are added. 
When coordinates and country are provided by reverse geocoding and coincide with those of \textsc{url}-based algorithm, if city and state lack, are added. 
On the contrary, the coordinates are included if cities of both geocoding methods overlap. 
Finally, affiliations are further enriched with foundation dates, entities, and acronyms. 
The result of the entire enriching procedure allows to obtain a dataset of Affiliations described by coordinates (from MAKG), a standardised city with coordinates, a state, a postcode, the standardised country with \textsc{iso} Alpha codes and official name\footnote{eventually plus the \textsc{iso} Alpha 2 code of the second country the affiliation could be localised.}, entity type, the foundation date, and \textsc{url}s. 


\section{Authors' Careers, Stocks and Flows}
\label{sec:careers_stocks_flows}

Authors' careers are computed by collecting their papers tagged with the year of publication. 
However, the MAKG makes it impossible to retrieve information on the authors' location. It doesn't include the relationship between publications, authors, and institutions (for which geographic information is available).
The relationship between the three entities is gathered from the latest available corresponding subset in MAG\footnote{Since the time coverage of datasets is different, i.e., MAKG 2020 while MAG 2019, this matching phase imposes a data loss.}. 
The MAG models the triple \textit{(paper, author, institution)} as ``has authors'' relationship with a direct edge from a publication to each of its authors \cite{wang2020microsoft}. 
Thus, careers and geolocated affiliations (Section \ref{subsec:reverse_geocoding}) are crossed to provide a set of paper and author pairs yearly linked to a geolocalised affiliation, i.e., \textit{(paper, author): year, affiliation\_ID, affiliation\_alpha2}.
The obtained data describes the authors' geolocalised careers with annual resolution. 
Starting from the hypothesis that the authors live where they are affiliated, thus in the country of their affiliations, it is possible to  define:
\begin{itemize}
    \item the author's \textit{annual location}, as the most frequent country among locations of institutions related to the author’s annual publications.
    \item the author's \textit{career nationality}, as the country of the first geolocated institution over the author's career.
\end{itemize}

By applying the existing literature to geolocalised careers, the concepts of \textit{i. working-nativity} (Theorem \ref{def:wn}), \textit{ii. authors' stock} (Theorem \ref{def:stock}) and \textit{iii. authors' flow}  (Theorem \ref{def:flow}) are introduced based on the definitions of \textit{migrants stocks} and \textit{migrants flows} \cite{desa2017handbook}. 
\begin{Theorem}
\label{def:wn}
An author A is a working-native of a country C if C is the author's career nationality. 
\end{Theorem}

\begin{Theorem}
\label{def:stock}
Authors' stock refers to the number of authors identified as international working migrants during a given year in a country. 
\end{Theorem}

\begin{Theorem}
\label{def:flow}
Authors' flow refers to the number of authors entering or leaving a given country in a year. 
\end{Theorem}

Following Theorems \ref{def:wn},\ref{def:stock} and \ref{def:flow}, the total number of authors based on countries is obtained by aggregating authors' annual locations. 
Then, stocks are computed by counting the number of non-working-natives researchers in each country. 
This is because an author cannot be defined as a working migrant of its working-native country. 
Flows are modelled as the direct graph between worldwide countries based on changes in authors' annual locations. 
A flow is represented by a weighted link between countries describing the origin and destination of a researcher's movement ($C_{Origin}$, $C_{Destination}$, respectively).
The weight is the number of authors who moved from ($C_{Origin}$) to ($C_{Destination}$). 
Moreover, flows are enriched with:
\begin{itemize}
    \item \textit{returners}: the number of authors located in the destination country for at least the second time during their career. 
    \item \textit{origin natives}: the number of authors leaving their working-native country. 
    \item \textit{destination natives}: the number of authors returning to their working-native country. 
\end{itemize}


\section{Networks of Collaboration}
\label{sec:networks}
The role of scientific collaborations, together with their topology and dynamics, have been extensively analysed in different research strands for several purposes~\cite{newman2001structure,montoya2018power,paraskevopoulos2020dynamics,hou2008structure}. 
Studies have been conducted at different resolution levels, e.g., micro-level (individuals), meso-level (institutions), and macro-level (countries) \cite{hou2008structure}.
According to the literature, two scholars are connected if they have authored a paper together~\cite{newman2001structure}. 
EMAKG provides the networks of collaborations by building authors' ego networks\footnote{An \textit{ego network} consists of a central node (\textit{ego}) and the nodes to which it is directly connected (\textit{alters}), plus the \textit{links} among the nodes.}. 
Various social relations can link together egos and alters depending on the network, e.g., working and personal relationships. 
In this case, an author's ego network is the weighted graph of its scientific collaborators in publishing papers.
To this end, an ad-hoc algorithm takes as input the relationships among papers, authors, and years of publications. Then, it computes the co-authors' list in each publication annually for each author. 
Further, since two authors may have published more than one paper, the links between nodes (co-authors) are tagged with the weight as the count of the shared publications.


\section{Papers Abstracts}
\label{sec:abstracts}
Papers abstracts are not included in the latest version of the MAKG. 
Abstracts have been extensively investigated, particularly in linguistics. 
Studies focus on type and provenance of publications~\cite{busch1995abstracts,amnuai2019analyses}, and research strands, e.g., medical~\cite{busch1995abstracts}, applied linguistics and educational~\cite{pho2008research,golebiowski2009prominent}, and biomedical~\cite{nam2016structuralizing} to analyse styles~\cite{guerini2012linguistic}, linguistic complexity~\cite{whissell1999linguistic} and rhetorical forms~\cite{tanko2017literary}. 

A semi-supervised pipeline is built to add linguistic data to abstracts obtained from the penultimate version of the MAKG. 
The method first infers the language code by using the LangDetect\footnote{LangDetect: \url{https://pypi.org/project/langdetect/}.}. 
Then, it employs Spacy\footnote{Spacy: \url{https://spacy.io/}.} and Html\footnote{Html: \url{https://pypi.org/project/html/}.} to clean and extract tokens, lemmas, and types from abstracts. 
The method provides texts of abstracts together with its \textsc{iso} 639-1 code, the list of tokens with frequency counts, and the list of types. 


\section{H-Index}
\label{sec:hin}
The h-index~\cite{hirsch2005index} (also known as Hirsch index/number) is a measure of the author's scientific achievements that considers both the number of papers published and the citations those receive. 
Despite its drawbacks\footnote{According tor~\cite{hirsch2020superconductivity}, the metric doesn't take into account research that deviates from the mainstream.}~\cite{cameron2005trends,favaloro2009journal}, it has become one of the most well-known metrics in academia. 
The index refers to the highest number \emph{h} such that an author has \emph{h} publications with at least \emph{h} citations. 
Google Scholar\footnote{\url{https://scholar.google.com/}.} is one of the most well-known academic search engines, but it has no official APIs. 
Among available 3rd party APIs, ScraperAPI\footnote{\url{https://www.scraperapi.com/}.} to be combined to prebuilt Google Scholar scraping libraries, e.g., Scholarly\footnote{\url{https://scholarly.readthedocs.io}.}; SERP API\footnote{\url{https://serpapi.com/}} and Publish or Perish\footnote{\url{https://harzing.com/resources/publish-or-perish}} are specifically designed for Google Scholar. 
However, Publish or Perish may requires proxy solutions, e.g., ScraperAPI; SerpWow\footnote{\url{https://www.serpwow.com/google-scholar-api}.} has no dedicated Google Scholar documentation; Scale SERP\footnote{\url{https://www.scaleserp.com/}.} cannot be customised and provides low-granularity data compared to most other APIs. 
All the 3rd party APIs are under subscription with different requests limits and pricing. 
Free available libraries include Scholarly, which allows searching authors by name, by the id in the url of an author's profile (in Google Scholar), by keywords, and by (titles of) publications, and Scholar.py\footnote{\url{https://github.com/ckreibich/scholar.py}.} which allows searching authors by name and by keywords. 
However, a) these libraries permit a few queries compared to dataset dimension, b) may return non-disambiguated results since searches are based on name, c) could not be consistent with the number of publications and citations per paper in the dataset. 
To face these limitations, the h-index is calculated directly from the data. 
The array of citations is computed starting from authors' publications. Then, the index is computed following three different methods, and results are cross-checked to ensure their reliability due to the lack of ground truth. 
I compute h-index by applying the Scholarmetrics\footnote{Scholarmetrics: \url{https://scholarmetrics.readthedocs.io/en/latest/index.html}.}, the function\footnote{\url{https://gist.github.com/jainsourabh/a3ee68f63632b55e5ada6c6a25e8620a/}.} derived by~\cite{hirsch2005index}, and the \textit{h index array} function\footnote{\url{https://gist.github.com/jainsourabh/506fed1a7ae327672b6f53c60be9ebff/}.}. 
The method achieves 100\% agreement over the three methods, providing an h-index for all authors.


\begin{figure}[b]
\centering
    \includegraphics[scale=.32]{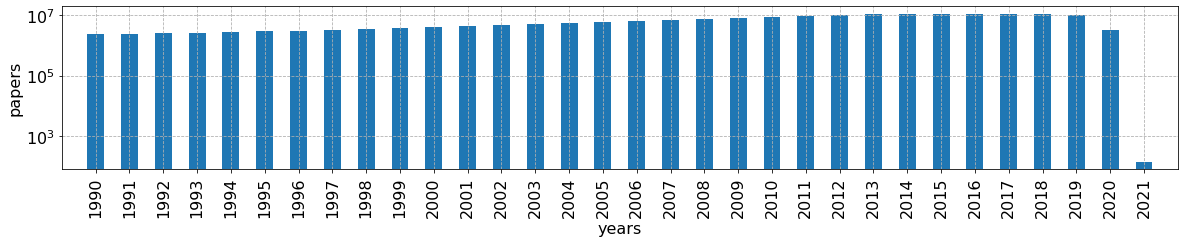}
    \caption{Distribution of the yearly number of publications (from 1990).}
    \label{fig:paper_distribution}
\end{figure}


\section{Fields of Study}
\label{sec:fos}

Fields of study represent research strands and concepts papers are associated with.  Information about which field(s) of study does a publication belongs to is very valuable for many tasks, but this may be often complicated to collect or calculate~\cite{herrmannova2016analysis}.
Each field of study is represented by a name, paper and citation counts, and hierarchy level of abstraction ranging in $[0,5]$.
Field studies are structured according to parent-child relationships, and each research field can have multiple parents. 
The latest MAKG version includes a descriptive classification obtained by assigning abstracts of papers to the 19 levels 0 top-fields of study from MAG (Section~\ref{sec:makg}). 
Together with the data loss~\cite{farberenhancing}, the classification method seems to be more suitable for research fields with a highly specialised and domain-dependent lexicon, i.e., geology, psychology, medicine, and biology. 
The result of descriptive classification is a list of tags describing the topic of the paper. 
Here, levels of abstraction with kinship relationships are used to propagate and assign \textsc{fos}s to fields. 
The 19 top-level \textsc{fos}s are directly tagged with the corresponding research area since these have no \textsc{fos} parents. 
Conversely, \textsc{fos} of lower levels (from 1 to 5), first, inherit the parents' tags. 
Then scores (as \textit{proportion}) in $[0, 1]$ ranges for each research area are computed based on the tags lists\footnote{The 19 top-level \textsc{fos}s have a 1.0 score.}. 
The \textsc{fos} labelling allows, in turn, to assign one or more research areas publications. 
For each paper, $a)$ is obtained the list of \textsc{fos}s to which it is associated; $b)$ are added up the scores by research areas; $c)$ the obtained scores are divided by the sum of the scores of all the research areas rescaling the score in the range $ [0,1]$. 


\begin{table}[t]
    \centering
    \caption{Distribution of type of publications}
    \begin{tabular}{l|r|l|l}
    \hline
    \textbf{Paper type} & \textbf{\# documents} & \textbf{\% over tot} & \textbf{\% over types}\\
    \hline
    Journal & 85,759,950 & 35.93 & 57.87\\
    Patent & 52,873,589 & 22.15 & 35.67\\
    Conference (in) & 4,702,268 & 1.97 & 3.17\\
    Book Chapter & 2,713,052 & 0.89 & 1.44\\
    Book & 2,143,939 & 1.13 & 1.83\\
    \end{tabular}
    \label{tab:papers}
\end{table}

\begin{figure}[b]
\centering
  \centering
  \includegraphics[scale=.45]{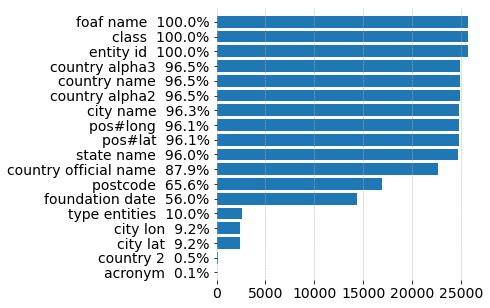}
    \caption{Features in the Affiliation subset}
    \label{fig:aff}
\end{figure}


\section{Data and Statistics}
\label{sec:stats}
The Enhanced Microsoft Academic Knowledge Graph (Appendix \ref{sec:appendice}) is a large dataset of scientific publications composed of several subsets representing entities and involved in publications and their relationships.  

\textbf{Papers. }Papers represent the core of the EMAKG graph. 
The dataset comprises 238,670,900 papers published from 1800 to 2021 with different rates. 
As shown in Figure \ref{fig:paper_distribution}, starting from 1900 the number of publications constantly grows\footnote{Note that the decrease in 2019, 2020, and 2021 is due to a gap between data collection (before June 2020) and data release.}. 
Publications are described by several properties, including the unique identifier, the entity class, and the unique identifier of the Journal, the conference series, and the conference instance in which the article is published. 
Also, papers have a rank, a family Id, and the counts about citations and references.
Other properties are based on DBpedia, DCMI, FaBiO, and PRISM data.
As shown by Table~\ref{tab:papers}, the most represented document type is journal articles which describe 35.93\% of total papers and 57.87\% of papers with a non-null type. 
Patents cover 22\% of the entire dataset, while conference papers, book and books chapters do not reach 2\% each, while 37.90\% of the total publications do not have a document type\footnote{This lacking is already underlined in~\cite{farberenhancing}.}. 
Papers are associated with abstracts for which are provided original texts, ISO639-1 codes, tokens with frequencies, and types (Section \ref{sec:abstracts}).


\begin{figure}[t]
  \centering
  \includegraphics[scale=.28]{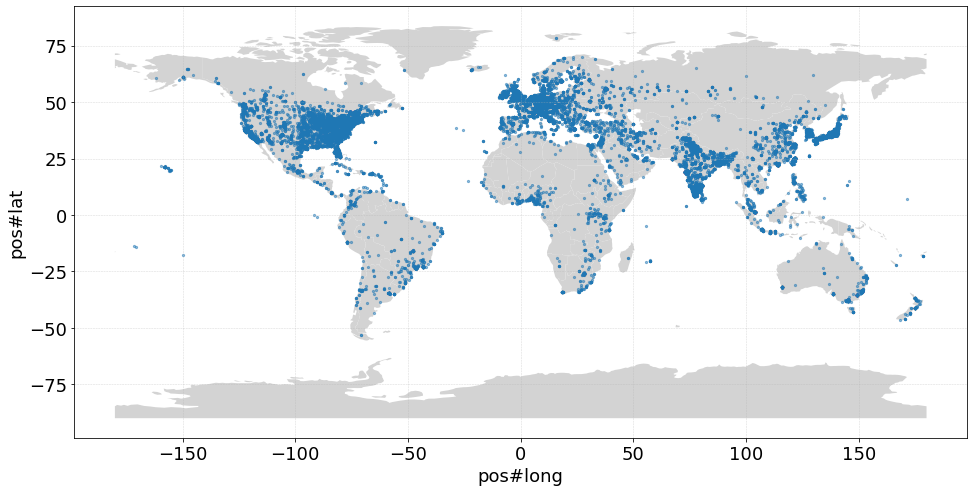}
    \caption{Locations of individual affiliations}
    \label{fig:aff_loc}
\end{figure}

\begin{figure}[b]
\centering
    \includegraphics[scale=.45]{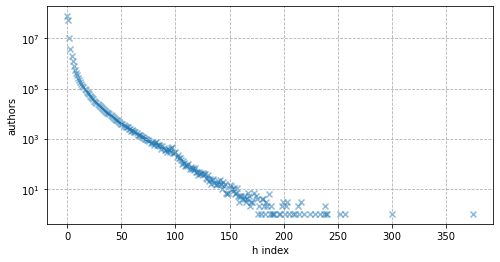}
    \caption{Distribution of authors' h-index}
    \label{fig:hind}
\end{figure}


\textbf{Affiliations. }Among the entities, the dataset comprises 25,768 enriched affiliations (Section \ref{sec:Geolocating_affiliations}) described by coordinates (latitude and longitude, from MAKG), a standardised city and its coordinates, a state, a postcode, the standardised country with \textsc{iso} Alpha codes and official name, and the \textsc{iso} Alpha 2 code of the second country the affiliation could localise in. 
Figure \ref{fig:aff} describes the features obtained together with the percentages with respect to the entire subset.
Most affiliations (about 96\%) is described by geographic information, e.g., coordinates, country name and \textsc{iso} Alpha codes.
Additionally, Figure \ref{fig:aff_loc} shows the location of individual affiliations based on geographic coordinates.
The world map is highly heterogeneous, with densely populated areas, e.g., United States of America and Europe
contrasted with areas with poor geolocation, e.g., African states and Russia. 
North America, Europe (especially Central), Brazil and Mexico, India, China, and Oceania include most affiliations. In contrast, Central America, western South America, Africa (excluding South Africa and Nigeria), Arab and Western Asian states, plus Russia are poorly represented.


\begin{figure}[t]
    \includegraphics[scale=.32]{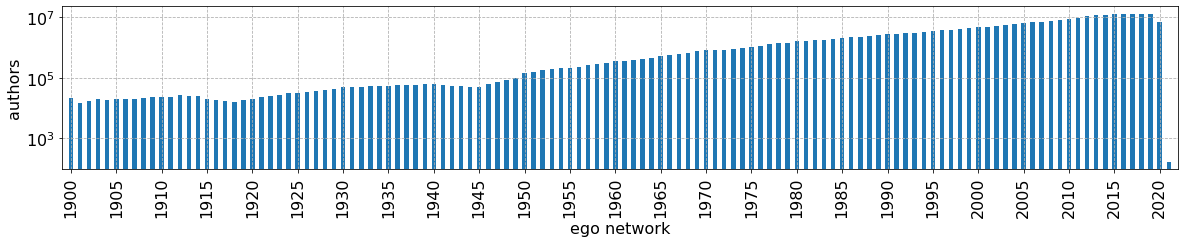}
    \caption{Distribution of authors over annual ego networks (from 1900)}
    \label{fig:egonet}
\end{figure}

\begin{figure}[b]
\centering
\begin{minipage}{.48\textwidth}
  \centering
  \includegraphics[width=.9\linewidth]{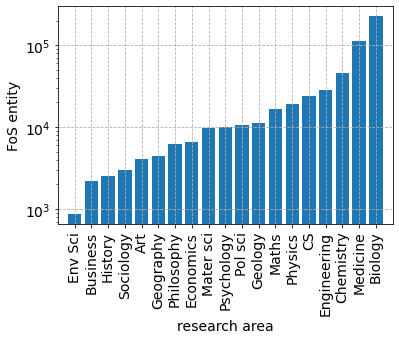}
  \captionof{figure}{Distribution of research areas along \textsc{fos}.}
  \label{fig:researchArea-dist}
\end{minipage}
\hfill
\centering
\begin{minipage}{.48\textwidth}
\includegraphics[width=.9\linewidth]{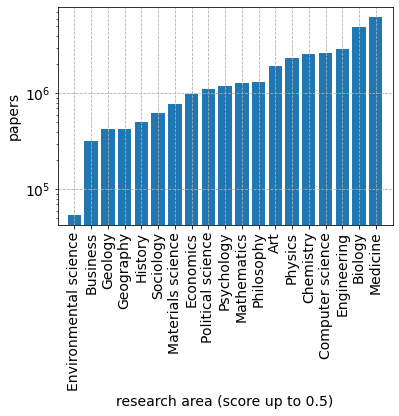}
  \caption{Distribution of research areas over papers (scores greater than or equal to 0.5.}
  \label{fig:paper_perArea_over05}
  \end{minipage}

\end{figure}


\textbf{Authors. }As MAKG, the EMAKG includes just over 243 million authors and more than 151 million disambiguated authors~\cite{farberenhancing}. An author is identified by a unique identifier and described by the class, the rank, the last known affiliation in MAKG, FaBiO name, and paper, paper family, and citation counts. 
Careers of authors\footnote{Careers are provided in .json format.} are computed following the method in Section \ref{sec:careers_stocks_flows} and include the publications and the related affiliations annually. 
On average, an author published 4.21 papers. 
Authors are further described using their h-index (Section \ref{sec:hin}). 
On average, an author has a 1.01 h-index, as in Figure \ref{fig:hind}. 
The five higher ones in the dataset reach respectively 240, 252, 257, 300, and 375, which might be partially plausible, but is likely misleading due to unclean data to some extent, as already underlined in some cases in~\cite{farberenhancing}. 
Information on authors include also their annual locations\footnote{Annual locations are released in .json format.} together with the affiliation and ego networks as the set of yearly co-authors. 
The time coverage of networks ranges from 1856 and 2018. The distribution of authors from 1900 is in Figure \ref{fig:egonet}.
Geolocalisation of 27,447,988 authors is provided from 1800 to 2020 and covers 195 worldwide countries.

\textbf{Field of Study. }
The dataset comprises more than 740 thousand \textsc{fos}s labelled with at least a field of study, thus the research area(s), to which the \textsc{fos} bellows according to the multiple parent-child relationships.
Most of the \textsc{fos} are labeled with one or two main research areas, while fewer than 4500 \textsc{fos} are tagged with more than six areas. 
Most of the \textsc{fos} refer to the so-called STEM\footnote{The acronym STEM stands for Science, Technology, Engineering, and Mathematics.} disciplines, thus science, technology, engineering and mathematics and any subjects that fall under these four disciplines as computer science (CS), biology, and chemistry. 
Conversely, humanities-related disciplines such as history, art, and philosophy seem characterised by fewer \textsc{fos}, as shown in Figure~\ref{fig:researchArea-dist}. 
By using \textsc{fos}, more than 44 million papers are tagged with at least a research area. 
Figure~\ref{fig:paper_perArea_over05} refers to distribution of research areas over papers obtained by summing scores (greater than or equal to $0.5$) of papers by research areas.

\begin{figure}[t]
\centering
\subfloat[]{\includegraphics[width=.7\linewidth]{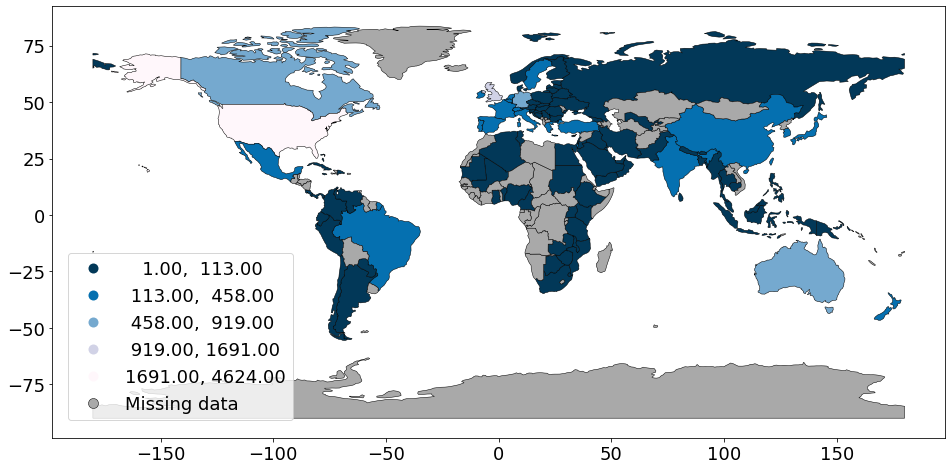}}\quad
\subfloat[]{\includegraphics[width=.7\linewidth]{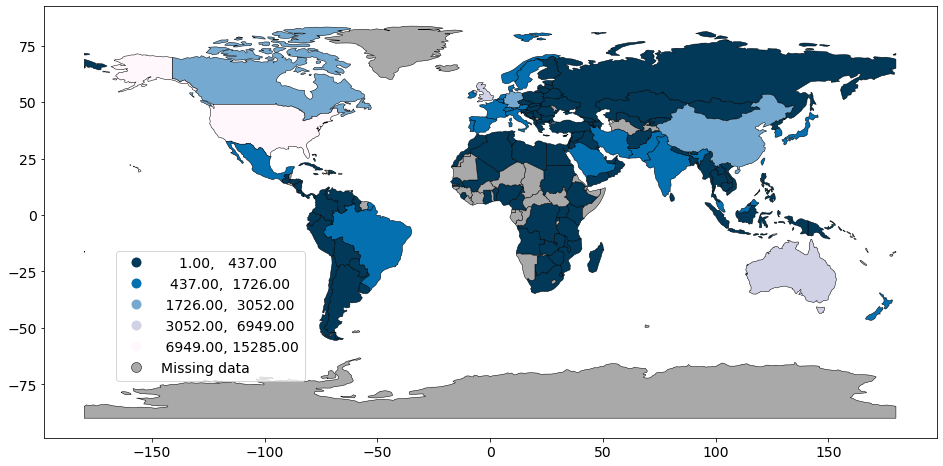}}
\caption{Worldwide stocks: (a) 2000, (b) 2019.}
\label{fig:stocks}
\end{figure}

\begin{figure}[t]
\centering
\subfloat[]{\includegraphics[width=.5\linewidth]{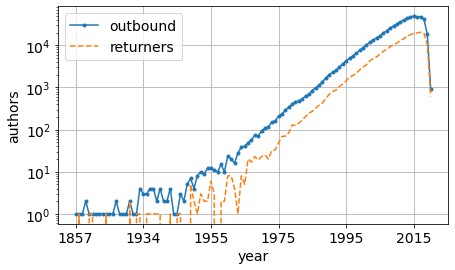}}\quad
\subfloat[]{\includegraphics[width=.5\linewidth]{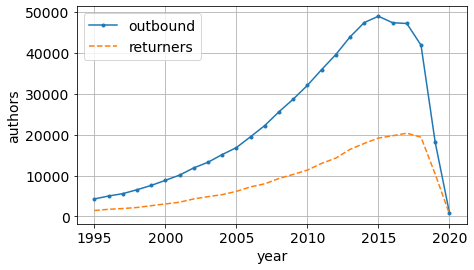}}
\caption{Total flows: (a) Entire time span (1857-2020), and (b) Focus up to 1995}
\label{fig:flowstrends}
\end{figure}

\textbf{Careers, Stocks  \&  Flows. }The EMAKG includes authors' careers as the set of their annual publications and related affiliations. 
Authors' careers are calculated by leveraging geolocalised affiliations and data about publications (Section~\ref{sec:careers_stocks_flows}). 
EMAKG provides careers for 27,647,403 authors. 
Geocalised careers are 27,448,058 (-199,345). 
The dataset also includes researchers' stocks (Theorem~\ref{def:stock}) and flows (Theorem~\ref{def:flow}) (Section~\ref{sec:careers_stocks_flows}) with annual temporal resolution and worldwide coverage. 
Both span from 1857 to 2020 with some sparse gap before 1945. 
Figure~\ref{fig:stocks} shows researchers' stocks in 2000 and 2019, respectively, and the general increasing number of authors worldwide and outline the consolidation of the power of some countries, e.g., the United States of America and the United Kingdom. 
Data allows studying different levels of spatial granularity annually, i.e., country level, continent level, and customised and ad-hoc geographical sets. 
Flows are represented as annuals direct graphs representing researchers' flows between country pairs by leveraging geolocalisation on affiliations and aggregating data. 
Figure \ref{fig:flowstrends} shows the trend of the movements of researchers along the entire time axis as well as in detail from 1995 onwards.
The global flows of researchers generally tend to grow over time with minimal decreases only before 1975.
From 2018 the trend is reversed.
This can be due to the data loss during merging sources with different time coverage.

\section{Application Scenarios}
\label{sec:usage}
The Enhanced MAKG is built on top of the Microsoft Academic Knowledge Graph.  Although significant additions have been made, most of the original structure has been conserved, including main original entities relationships, and properties. 
Thus, the EMAKG could be exploited in all the uses cases and applications of MAKG. 
Besides, the EMAKG $a)$ reintroduces no longer available subsets, e.g., abstracts, $b)$ merges new knowledge retrieved from external resources (i.e., Wikipedia) and libraries, and $c)$ adds new relationships among entities computed by aggregating data of the dataset. 
The proposed improvements could open new analyses and applications of the dataset. 
Among the possible research areas, data could be used to measure institutions' research output~\cite{tzitzikas2020can} and for the science of science~\cite{fortunato2018science} also considering the geographical dimension at different spatial resolutions. 
Authors' connections can be leveraged to study the knowledge exchange. 
Moreover, researchers' flows and stocks can be compared with official statistics to study highly skilled mobility and migration. 
Linguistics and computational linguistics could benefit from abstracts and pre-computed tokens and types for language studies.
Also, the annual authors' ego network can be explored for research in network analysis. However, thanks to the amount and multifaceted nature of information in the Enhanced MAKG and the easy access, e.g., .csv and .txt, data can be analysed in several fields and for multiple purposes.


\section{Discussion and Conclusions}
\label{sec:concl}
This article presents the Enhanced MAKG, an enriched version of the Microsoft Academic Knowledge Graph, and methods developed to build it. 
The main aim of the dataset is $a)$ making data accessible and easy to use and $b)$ enriching available information to allow and facilitate new analysis. 
Further, a set of methods aiming to retrieve, standardise, and add new information about existing entities are developed to improve available data.
Geographical information and geolocalisation are enhanced by combining reverse geocoding, information retrieval, and data integration. 
Authors-related data includes working connections (ego networks) and movements between institutions, publications, and general information. 
Further, EMAKG provides authors' annual locations and career nationalities, together with worldwide yearly stocks and flows. 
Among others, the subsets include $a)$ fields of study (and publications) labelled by their discipline(s); $b)$ abstracts and linguistic features, i.e., standard language codes, tokens, and types; $c)$ entities' general information, e.g., date of foundation and type of institutions; and $d)$ academia related metrics, i.e., h-index.
The resulting dataset maintains all the characteristics of the parent datasets and includes a set of additional subsets and data that can be used for new case studies relating to network analysis, knowledge exchange, linguistics and computational linguistics, and mobility and human migration, among others.

\vspace{6pt}

\acknowledgments{This work is supported by the European Union – Horizon 2020 Program under the scheme “INFRAIA-01-2018-2019 – Integrating Activities for Advanced Communities”, Grant Agreement n.871042, “SoBigData++: European Integrated Infrastructure for Social Mining and Big Data Analytics”, and by  the  Horizon2020  European  projects  ``HumMingBird – Enhanced migration measures from a multidimensional perspective'', Grant Agreement n. 870661. }

\conflictsofinterest{The author declare no conflict of interest. 
The funders had no role in the design of the study; in the collection, analyses, or interpretation of data; in the writing of the manuscript, or in the decision to publish the~results. } 




\appendixtitles{yes} 
\appendixstart
\appendix
\section[\appendixname~\thesection]{Enhanced Microsoft Academic Graph}
\label{sec:appendice}
\textbf{EMAKG version 0.0}. Version 0.0 provides a set of EMAKG subsets, some of which are in abridged form:

\begin{itemize}
\item \textit{01.AffiliationsGeo}: Affiliations subset.
\item \textit{03.ConferenceInstance}s: Conferences subset. 
\item \textit{04.Conference Series}: ConferenceSeries subset.  
\item \textit{05.Journals}: Journals subset.  
\item \textit{06.24.PaperAuthorAffiliations\_Disambiguated}: Relationships between papers and disambiguated authors.  
\item \textit{09.PaperResources}: \textsc{URL}s and resources of publications.
\item \textit{10.Papers}: Papers subset. 
\item \textit{12.EntityRelatedEntities}: Connections between entities.
\item \textit{13.FieldOfStudyChildren}: Field of study kinship relations. 
\item \textit{14.FieldOfStudyExtendedAttributes}: Fields of study co-references between different datasets.
\item \textit{15.FieldsOfStudy}: Fields of study subset.
\item \textit{16.PaperFieldsOfStudy}: Relationships between papers and fields of study. 
\item \textit{18.RelatedFieldOfStudy}: Relationships between symptoms, medical treatments, disease causes and fields of study.
\item \textit{19.PaperCitationContexts}: Contexts of citations in CiTO.
\item \textit{20.AbstractsProcessed\_Chunk0-14}: Chunk of processed abstracts. 
\item \textit{22.FieldOfStudyLabeled}: Tags and scores of fields of studies.
\item \textit{23.Authors\_disambiguated}: Disambiguated authors subset. 
\item \textit{24.PaperAuthorAffiliation\_Disambiguated}: Relationships between papers, disambiguated authors and affiliations. 
\item \textit{25.AuthorORCID}: Authors' ORCIDs. 
\item \textit{26.AuthorCareer}: Authors' yearly publications. 
\item \textit{27.AuthorYearLocation}: Authors' yearly locations. 
\item \textit{28.AuthorEgoNetworks\_2000-2014}: Authors' ego networks from 2000 to 2014. 
\item \textit{29.CountryAnnualFlowsAggregated}: Flows aggregated by country and year.
\item \textit{30.FlowsAnnual}: Annual country to country flows.
\item \textit{31.StocksAnnual}: Annual stocks aggregated by country.
\item \textit{32.PaperFieldsOfStudyLabeled}: Publications tagged with fields of studies.
\item \textit{33.Authors\_disambiguated\_Hindex}: Authors' Hindex.
\end{itemize} 

\textbf{EMAKG version 1.0}. In addition to the subsets of version 0.0, version 1.0 also includes:
\begin{itemize}
\item \textit{02.Authors}: Authors' subset.
\item \textit{06.PaperAuthorAffiliations}: Relationships between papers and authors.  
\item \textit{07.PaperExtendedAttributes}: Patent numbers and PubMedIds.
\item \textit{08.PaperReferences}: References in CiTO.
\end{itemize} 

Furthermore, version 1.0 provides the full version of some subsets that are released in abridged form in version 0.0:
\begin{itemize}
\item \textit{20.AbstractsProcessed}: Processed abstracts. 
\item \textit{28.AuthorEgoNetworks}: Authors' ego networks.
\end{itemize}

\begin{adjustwidth}{-\extralength}{0cm}

\reftitle{References}



\bibliography{main}

%


\end{adjustwidth}
\end{document}